\documentstyle[preprint,aps,epsf]{revtex}
\begin{document}

\draft

\title{Quantum superconductor-metal transition}

\author{B.Spivak}

\address{Physics Department, University of Washington, Seattle, WA 98195}

\author{A.Zyuzin}

\address{A.F.Ioffe Physical- Technical Institute, 194021, St.Petersburg, 
Russia}

\author{M.Hruska}

\address{Physics Department, University of Washington, Seattle, WA 98195}

\maketitle

\begin{abstract}
We consider a system of superconducting grains embedded in a normal metal.
At zero temperature this system exhibits a quantum
superconductor-normal metal phase transition. This transition can take place at
arbitrary large conductance of the normal metal.   
\end{abstract}


\newpage

\section{Introduction}

Quantum superconductor-insulator (or superconductor-normal metal)
phase transitions can take place at zero temperature due to variation
 of parameters of the system.
 For example in experiments \cite{orr,val} the transition takes place as a
function of the degree of disorder in a superconducting film. In experiments of
 \cite{heb,jval,kapitulnik} the transition was mediated by a magnetic
field.
 
It has been suggested that a disordered superconducting film can be 
described as a network of Josephson-coupled superconducting grains
 shunted by resistors \cite{Efetov,simanek,sudip,vic}. 
In this case the Coulomb interaction between electrons in grains 
suppresses fluctuations of the number of electrons in a grain and, due to
uncertanty
principle, increases the amplitude of fluctuations of the phase 
of the superconducting order parameter \cite{simanek,sudip,girvin,vic}.
The competition between the charging energy and 
the Josephson inter-grain coupling energy leads to the phase transition. 
Shunting resistors play a double role
in the model: a) dissipation in the resistors tends to suppress 
 fluctuations of the phase of the 
order parameter b)tunneling between superconducting grains and the
 resistors renormalizes the capacitance of the grains and thereby
 the charging energy.
As a result, in the framework of this model and in two dimensional case
 the onset of superconductivity 
corresponds to the normal state film's conductance $G$ of the film of order
of
$\frac{e^{2}}{h}$.
Since at $G<\frac{e^{2}}{h}$ the system should be 
in the insulating state, it has been conjectured that the transition is
of the superconductor-insulator type \cite{mfisher,sudip,girvin,vic} and 
that the behaviour of the system near the transition is universal. 
On the other hand, a 
renormalization  group analysis \cite{finkelshtein}, which starts from
 a perturbation theory for slightly disordered uniform
 superconductors, showed in 2D case a zero temperature
superconductor-metal
transition.  

Recently Feigelman and Larkin \cite{feigelman} reconsidered the problem 
in the framework of a model of superconducting grains
 embedded in a normal metal film. They showed that a) the
superconductor-normal metal transition takes place and b) deep in
the metallic phase parameters of the system can
be calculated with the help of perturbation theory.  At
$G>\frac{e^{2}}{h}$, however, the critical concentration of grains
 turned out to be exponentially small.

On the other hand, the theories concerning granular superconductors
\cite{Efetov,simanek,sudip,vic}
were done in the limit when the modulus of the order parameter
 on a grain does not fluctuate. In the case of no
 reflection on the superconductor-normal metal boundary this corresponds
 to $R\gg \xi$. Here $R$ is the
 grain's radius, $\xi=min[\frac{v_{F}}{\Delta_{0}};
\sqrt{\frac{D}{\Delta_{0}}}]$ is the zero temperature coherence length
of the bulk superconductor, $D=\frac{v_{F}l}{3}$ is the electron diffusion 
coefficient in the metal, $v_{F}$ is the Fermi velocity,
$l$ is the electron elastic mean free path and $\Delta_{0}$ is the zero
 temperature value of the gap in the bulk superconductor.
 
In this paper we consider the opposite case when $R<\xi$ and show that  
the  zero temperature quantum superconductor-metal 
transition as a function of grains' concentration can take place in samples
with
arbitrary large conductance
 and can exist even in the absence of any disorder in the sample and 
in the approximation when electrons in the normal metal do not interact.
The critical concentration of the grains turns out to be relatively
large. 
 
We will consider a quasi-two-dimensional film of thickness 
$a\ll \xi$, which 
consists of superconducting grains of radius $R$ embedded
in
a nonsuperconducting metal. The results can be easily generalized for d=3 case.
 We will assume that densities of states $\nu$ of the
superconductor and the metal are the same and that the spatial
dependence of the electron-electron interaction constant in the
 the sample has the form
\begin{equation}
\lambda(\bbox{r})=\left\{
 \begin{array}{ll}
\lambda_{s}>0 & \mbox{if $|\bbox{r}-\bbox{r}_{i}|<R$}
\\
\lambda_{N}<0 & \mbox{if $|\bbox{r}-\bbox{r}_{i}|>R$} \end{array} \right. \
 \end{equation}
Here $\lambda_{s}$ and $\lambda_{N}$ are electron interaction constants
 in the superconductor and in the metal respectively; $\bbox{r}_{i}$
are coordinates of centers of the superconducting grains
 and index $i$ labels the grains.
At zero temperature the linearized mean field equation \cite{abr} for the
order
 parameter $\Delta(\bbox{r})$ 
\begin{equation}
\Delta(\bbox{r})=\lambda(\bbox{r})\int
d\bbox{r}_{1}K(\bbox{r},\bbox{r}_{1})
\Delta(\bbox{r}_{1})
\end{equation}
has a solution $\Delta(\bbox{r})=f(|\bbox{r}-\bbox{r}_{i}|)$
at $R=R_{c}^{(mf)}$. Assuming no reflection at the superconductor-metal
boundary
we get $R_{c}^{(mf)}\sim\xi$ \cite{DeGennes}.
Here 
$K(\bbox{r},\bbox{r}_{1})=\int d \epsilon
G(-\epsilon,\bbox{r},\bbox{r}_{1})G(\epsilon,\bbox{r},\bbox{r}_{1})$,
$G(\bbox{r},\bbox{r}_{1})$ is the electron 
Green's function in the normal metal.
It is convenient to normalize $\int d \bbox{r} |f(\bbox{r})|^{2}=1$. 
At $R<R_{c}^{(mf)}$ the mean field value of the order parameter
is zero ($\Delta(\bbox{r})=0$). To find the value of $\Delta(\bbox{r})$
 in the case $R>R_{c}^{(mf)}$ one has to add to Eq.2 terms nonlinear
 in $\Delta$ \cite{abr}.

\section{Quantum fluctuations of the order parameter in an individual grain}

We will consider the case $R<R_{c}^{(mf)}$ when for an individual grain
Eq.2 does not have nonzero mean field solution.
To calculate the correlation function of quantum fluctuations of the order
 parameter it is convenient to use a  parametrization
$\Delta_{i}(\bbox{r},t)=\alpha_{i}(t)f(|\bbox{r}-\bbox{r}_{i}|)$, which
 reflects the fact that, at $(R^{(mf)}_{c}-R)\ll R^{(mf)}_{c}$, the
 amplitude of quantum fluctuations of
$\alpha_{i}(t)$ is large while the amplitude of fluctuations  of
 the shape of $f(r)$ is small. 
To describe dynamics of the order parameter in a grain  we use the 
effective
action 
\begin{equation}
S_{i}=\nu \tau_{0}\int \frac{d\omega}{2\pi} (-i|\omega|+
\frac{1}{\tau})|\alpha_{i}(\omega)|^{2}
\end{equation}
whose derivation we outline in the Appendix. 
Here $\tau_{0}=min[\frac{r}{v_{F}}; \frac{R^{2}}{D}]$ is the the
 time of electron flight through the grain and
\begin{equation}
\tau=\frac{\tau_{0}R}{R_{c}^{(mf)}-R}
 \end{equation}
 Using Eq.3 we get
\begin{equation}
<\alpha_{i}(\omega),\alpha^{*}_{i}(-\omega)>_{(0)}=
\frac{1}{\nu \tau_{0}(-i|\omega|+\frac{1}{\tau})}
\end{equation}
which in $t$ representation corresponds to
\begin{equation}
<\alpha_{i}(t),\alpha^{*}_{i}(0)>_{(0)}= \left\{\begin{array}{ll}
\frac{1}{\nu \tau_{0}}(\frac{\tau}{t})^{2}
 & \mbox{if $t\gg \tau$}
\\
\frac{1}{\nu \tau_{0}}
i[-i\pi+2ln(\frac{t}{\tau})]
 & \mbox{if $t\ll \tau$} 
 \end{array} \right. \
\end{equation}
Here the subscript $"(0)"$ indicates that the correlation function
is calculated for a single existing
grain $"i"$. Eqs.5,6 hold as long as terms nonlinear in $|\alpha_{i}|^{2}$
 in the effective action can be neglected,
i.e. if $\tau\ll \delta^{-1}$. Here $\delta=(\nu R^{2}a)^{-1}$ is the
 average level spacing in the grain.
They correspond to the casual Green's function
$G_{c}=<T(\Psi_{\sigma}(\bbox{r},t)
\Psi_{-\sigma}(\bbox{r},t)\Psi^{+}_{\sigma}(0,0)
\Psi^{+}_{-\sigma}(0,0))>$, where $\sigma$ is the spin index.
To get the retarded Green's function
 $G_{R}=<\theta(t)[\Psi^{+}_{\sigma}(\bbox{r},t)\Psi^{+}_{-\sigma}(\bbox{r},t),
\Psi_{\sigma}(0,0)\Psi_{-\sigma}(0,0)]_{+}>$
 one has to make analytical
 continuation of Eq.5 with respect to $\omega$ and then to make a
 Fourier transform. As a result, $G_{R}(t)\sim \exp(-\frac{t}{\tau})$.
It is interesting that the asymptotic time dependence of Eq.6 is the same as
the one obtained in the case $R\gg
R^{(mf)}_{c}$ \cite{feigelman} with the help of a complicated
renormalization group
 analysis of the Caldeira-Leggett effective action \cite{leg}. In the
latter case
there is a
 non-zero mean field
order parameter on a grain and the correlation function decays with time
due
 to phase fluctuations mediated by
the interaction with quantum electromagnetic fluctuations in conducting
environment.
At $G>\frac{e^{2}}{h}$, however, the corresponding correlation time turns out
 to be exponentially large, which is  different from the case
 $R<R^{(mf)}_{c}$ Eq.4.

\section{The superconductor-metal transition in a system of superconducting
 grains}

Let us consider now the case when the concentration of 
superconducting 
grains $N$ embedded into the quasi-two-dimensional metallic film is finite.
 To describe this system in the case when $\alpha_{i}$ are small we will use
the effective action $S=S_{0}+S_{int}$
where $S_{0}=\sum_{i}S_{i}$,
 while $S_{int}$ describes inter-grain interaction via the normal metal
and also the nonlinear in $|\alpha_{i}|^{2}$ contributions to the action.
\begin{equation}
S_{int}=-\sum_{i,j}\int d\omega
J_{ij}\alpha_{i}(\omega)\alpha^{*}_{j}(-\omega)
+b\sum_{i}\int d\omega_{1}d\omega_{2}
d\omega_{3}\alpha_{i}(\omega_{1})\alpha^{*}_{i}(\omega_{2})
\alpha_{i}(\omega_{3})\alpha^{*}_{i}(-\omega_{1}-\omega_{2}-\omega_{3})
\end{equation}
where $J_{ij}$ have the meaning of Josephson coupling between grains $i$ and
$j$, and $b\sim \frac{R^{2}_{c}\nu}{a D^{2}}$.

The mean field approximation corresponds to the minimum of $S$ at $\omega=0$.
If $|\bbox{r}_{i}-\bbox{r}_{j}|$ is small enough the solution of the Uzadel
equation Eq.25 yields Josephson coupling between two grains of the form
($|\bbox{r}_{i}-\bbox{r}_{j}|\gg R$) 
  \begin{equation}
J_{ij}=\frac{\nu R^{2}}{|\bbox{r}_{i}-\bbox{r}_{j}|^{2}}
\end{equation} 
 
However, in the case of finite two-dimensional concentration $N$ 
such an expression for $J_{ij}$ would lead to 
a logarithmic divergence of the ground state energy density.

On the other hand, at large $|\bbox{r}_{i}-\bbox{r}_{j}|$ electron-hole
pairs
diffusing through the metal between
 grains $i$
and $j$ will experience Andreev scattering from the superconducting grains 
situated between grains $i$ and $j$. An example of such a grain
$"k"$ is shown in Fig.1. 
Due to
the Andreev nature of reflection \cite{andreev} electrons scattered by the
grain
 $"k"$   
will be reflected into a hole moving 
in the direction opposite to the initial direction of electron's motion.
Thus this electron will never reach the grain $j$.
As a result we have 
\begin{equation}
J_{ij}=\frac{\nu R^{2}}{|\bbox{r}_{i}-\bbox{r}_{j}|^{2}}
\exp(-\frac{|\bbox{r}_{i}-\bbox{r}_{j}|}{L^{(mf)}_{0}})
\end{equation}
 
\begin{equation}
L^{(mf)}_{0}=(\frac{l}{NR}\frac{Ra^{1/2}\Delta_{0}}{\alpha_{(mf)}})^{1/2}
\end{equation}
Therefore the mean field equation for the ground state 
mean field value $\alpha_{(mf)}$ of the order parameter on a grain has a form
\begin{equation}
\frac{\nu \tau_{0}}{\tau}-2\pi\nu R^{2}N \ln(N^{1/2}
L^{(mf)}_{0})+b\alpha_{(mf)}^{2}=0
\end{equation}
Thus at $T=0$ an exponentially small mean field solution for the order 
parameter 
\begin{equation}
\alpha_{(mf)}\sim \Delta_{0}l a^{1/2}
 \exp(-\frac{\tau_{0}}{\tau R^{2}N})
\end{equation}
exists at arbitrary small $N$.

However, at small enough $N$ the amplitude of quantum fluctuations of the
order
parameter Eqs.5,6 becomes larger than its mean field value. In this
case 
the mean field theory is not applicable. 
 To show that at small $N$ quantum fluctuations 
destroy superconductivity completely and that the normal metal state is stable
we use a perturbation theory procedure similar to the one used in
\cite{feigelman}.
A requirement for the perturbation theory in therms of $J_{ij}$ to be
valid is the convergence
of 
the integral
\begin{equation}
\int <\alpha_{i}(t)\alpha^{*}_{i}(0)>dt < \infty
\end{equation}
which in our case follows from Eq.6.
The integral in this case equals $\frac{\tau}{\nu \tau_{0}}$.

Again, the Josephson couplings of the form Eq.8 would lead to divergency
of 
the perturbation theory.
To cut off the divergence in the absence of the magnetic field one can consider
the case 
\cite{feigelman} when 
there is a repulsion between electrons in the metal and $\lambda_{N}\neq 0$ .
 Then in the two-dimensional case
 we have $J_{ij}\sim \nu R^{2}r^{-2}[1+2\nu| \lambda_{N}|ln(\frac{r}{R})]^{-2}$
 and the perturbation theory will converge on the lengthscale 
$L_{\lambda}\sim R \exp(\frac{1}{\nu |\lambda_{N}|})$. 
In the presence of a weak magnetic field $H$
the Josephson intergrain coupling decays exponentially on distances larger than the 
magnetic length $L_{H}=\sqrt{\frac{m}{eH}}$.
Thus the cut off length relevant for the convergence of the
 perturbation theory with respect to the
term in Eq.7 proportional to $J_{ij}$ is $L_{0}=
\min[L_{\lambda};L_{H}]$. It gives small corrections to the
 Eqs.5, 6 as long as $NR^{2}ln(L_{0}N^{1/2})\ll
\frac{\tau_{0}}{\tau}$. 
In the opposite case the ground state of the system is superconducting.
Thus we can estimate a relation between the critical concentration
 of the grains $N_{c}$ and their critical radius $R_{c}$ from the equation
\begin{equation}
\frac{|R_{c}^{(mf)}-R_{c}|}{R_{c}^{(mf)}}\sim
 R^{2}_{c}N_{c} ln(L_{0}N^{1/2}_{c})
\end{equation}
For example, in the case $R<R_{c}^{(mf)}$ and at $H=0$ we have the estimate
\begin{equation}
N_{c}\sim \frac{1}{R^{2}_{c}} \nu|\lambda_{N}|
\end{equation}
This is different from the case $R\gg R_{c}^{(mf)}$ where at 
$G>\frac{e^{2}}{h}$ the critical concentration is
exponentially small. The difference originates 
from the difference in correlation times.

Neglecting the second term in Eq.7 and assuming for simplicity that 
superconducting grains form a square lattice we get
 an expression for the correlation functions 
\begin{eqnarray}
<\alpha_{k}(\omega)\alpha^{*}_{l}(-\omega)>=\frac{1}{\nu \tau_{0}}
(-i|\omega|+\frac{1}{\tau}+\frac{1}{\nu \tau_{0}}J_{ij})_{k,l}^{-1} \nonumber 
\\ =\frac{1}{\nu \tau_{0}}\sum_{\bbox{q}}\exp(-i\bbox{q}\bbox{r})
\frac{1}{-i|\omega|+\frac{1}{\tau}-K(\bbox{q})}
\end{eqnarray} 
where components of $\bbox{q}$ are whole multiples of $\frac{2\pi}{L}$
and $L$ is the samle size.
 \begin{equation}
K(q)=\frac{1}{\nu \tau_{0}}\sum_{i} exp(i\bbox{q}\bbox{r})K(\bbox{r}_{i})
\end{equation}
At $|\bbox{q}|N^{-1/2}\ll 1$ we have 
\begin{equation}
K(\bbox{q})=- \pi DN  ln((1+(qL_{0})^{2})L_{0}^{2}N)
\end{equation}
 The mean field Eqs.16-18 are valid because the characteristic radius of
 interaction between grains $N^{-1/2}ln(L_{0}N^{\frac{1}{2}})\gg N^{-1/2}$
 is much larger than the average intergrain distance.

Qualitatively the picture of quantum superconducting fluctuations in 
the normal metal is similar to superconducting fluctuations in
 uniform metals at temperatures which are close 
to the critical one \cite{asl,maki}:

a) Due to quantum fluctuations  conductivity of the system is enhanced
as compared to the normal metal value.
It exhibits a big positive magnetoresistance.

b) At small magnetic field the zero temperature
Hall coefficient
is suppressed as compared to its normal metal value \cite{fei}.
It also exhibits strong magnetic field dependence.

c) Diamagnetic susceptibility is enhanced and exhibits a
 strong nonlinearity as a function of magnetic field.

d) Energy dependence of the density of states at the 
Fermi surface has a dip, whose amplitude is magnetic field dependent.

\section{Conclusion}
We have considered a model in which superconducting grains with
the radius $R<R_c$ are embedded in the normal metal and have 
shown that 
in this case there is a zero temperature quantum superconductor-normal
metal phase transition as a function of $N$ and $R$. This
 transition is driven, primarily, by fluctuations of the modulus
 of the superconducting order parameter. Though the 
parameters of the transition, in principle, depend on 
$D$, it exists even in the case when there is no disorder
in the sample. 
The critical concentration of grains turns out to
be relatively large. 

Calculations presented above did not take into account localization
effects 
\cite{anderson,khm}. 
In d=3 case they are small. The 
question of whether in two-dimensional case the metallic state is
localized \cite{mfisher} requires additional investigation. 
On the other hand, as we have discussed, the transition can take place at
very large sample conductances when the localization length is also very large. 

We would like to mention that in the presence of the electron-electron 
repulsion in the metal
the metal-insulator transition 
can exist even in the mean field approximation described by the Eq.2. 
For example, in the case $\lambda_{s}\sim |\lambda_{N}|$ and
$R<R_{c}^{mf}$
 we have $N_{c}\approx \frac{1}{R^{2}}$.
  
Finally, we would like to mention, that in our opinion experimental data
on 2-D films \cite{orr,heb,val,jval,kapitulnik} do not contradict the
possibility of a zero temperature superconductor-normal metal quantum phase
transition. 
 
\section{Appendix}
We start with a standard  expression for the partition function
in a superconductors (see for example \cite{schoen}):
\begin{equation}
Z=\int D\Delta D\Delta^{*} \exp (iS_{eff})
\end{equation}
\begin{equation}
S_{eff}=\int d\bbox{r} dt[\frac{|\Delta|^{2}}{\lambda(\bbox{r})}+
\int_{0}^{1} d\eta Tr[\hat{\Delta}(\bbox{r},t)
\hat{G}_{\eta}(\bbox{r}t,\bbox{r}t)]]
\end{equation}
where $\hat{G}_{\eta}(\bbox{r}t,\bbox{r}'t')$ is a matrix Green function
which is
a solution of the Gorkov equation
\begin{equation}
[-i\frac{d}{dt}-\xi(i\nabla)\sigma_{z}+u(\bbox{r})-
\eta\hat{\Delta}(\bbox{r},t)]\hat{G}_{\eta}(\bbox{r},t;\bbox{r}'t')=
\delta(\bbox{r}-\bbox{r}')\delta(t-t')\hat{I}
\end{equation}
Here $u(\bbox{r})$ is the external potential. 
\begin{equation}
\hat{\Delta}=\left( \begin{array}{cc}
0 & \Delta
\\
\Delta^{*} & 0 \end{array} \right )
\end{equation}
and $\hat{I}$ is a $2*2$ unit matrix in Nambu space. We assume that $u(\bbox{r})$
 has a  white
 noise statistics with correlation functions  $<<u(\bbox{r})>>$ and 
$<<u(\bbox{r})u(\bbox{r}')>>=
\frac{v_{F}}{l\nu}\delta(\bbox{r}-\bbox{r}')$. Here brackets $<< >>$
 stand for averaging over realizations of the scattering potential.

Averaging Eq.21 over realizations of $u(\bbox{r})$, 
neglecting all weak localization and mesoscopic corrections, 
and making the diffusion approximation we get Uzadel equations
for the normal and the anomalous Green's functions (See, for example,
\cite{ovch})

\begin{equation}
\omega F(\omega,\bbox{r})+\frac{1}{2}D(F(\omega,\bbox{r})
\nabla^{2}G(\omega,\bbox{r})-G(\omega,\bbox{r})
\nabla^{2}F(\omega,\bbox{r}))=\eta\Delta^{*}(\bbox{r})G(\omega,\bbox{r})
\end{equation}
\begin{equation}
G^{2}+|F|^{2}=1
\end{equation}

Expanding Eqs.20-24 with respect to $\Delta$ we get
an expression for the effective action Eq.3 for an individual grain.

To get Eqs.9,10 we have to solve the Uzadel equation Eq.24
 in the normal metal between the superconducting grains. 
This solution corresponds to an electron diffusion in the metal 
and Andreev reflections from the superconducting grains. 
In the case $R\gg R_{c}^{mf}$ we have $|\Delta(t)|=\Delta_{0}$ 
and the Andreev scattering cross section equals to $R^{2}$.
Since in our case $\Delta(t)\ll \Delta_{0}$ the cross section
 is of order of $R^{2}\frac{\Delta^{2}(t)}{\Delta_{0}^{2}}$
and we get Eq.9,10. 

This work was supported by Division of Material Sciences, 
U.S.National Science Foundation under Contract No. DMR-9205144.
We would like to thank A.Andreev, S.Chakravarty, M.Feigelman,
 A.I.Larkin, D.Khmelnitskii, S.Kivelson, and F.Zhou for 
useful discussions.

\newpage

\begin{figure}
  \centerline{\epsfxsize=10cm \epsfbox{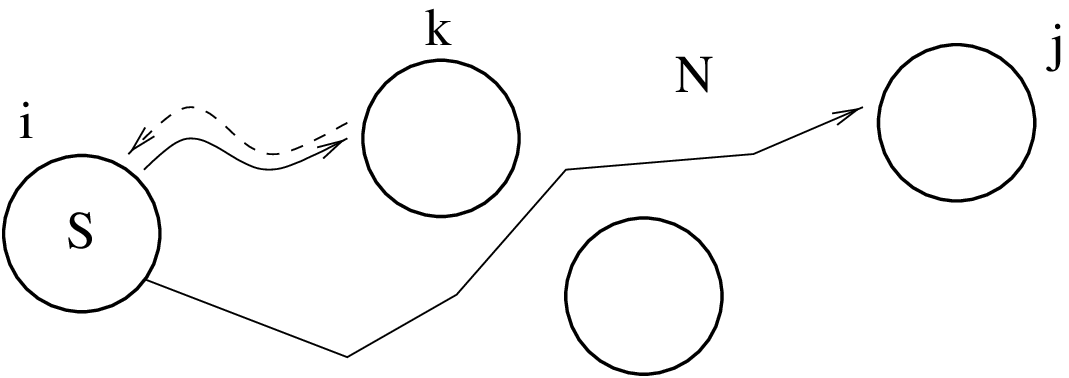}}
  \caption{A system of superconducting grains embedded into a normal
metal. $S$ stands for the superconductor while $N$ stands for the
normal metal. Solid lines correspond to trajectories of electrons
diffusing between grains $i$ and $j$. The dashed line corresponds to
the hole reflected from the grain $k$.} \
  \label{fig:fig1}
\end{figure}

\end{document}